\documentclass[12pt]{iopart}
\usepackage{iopams}
\begin{document}

\title{Are Maxwell's equations Lorentz-covariant?}
\author{D V Red\v zi\' c}

\address{Faculty of Physics, University of Belgrade, PO
Box 44, 11000 Beograd, Serbia} \eads{\mailto{redzic@ff.bg.ac.rs}}

\begin{abstract}
The statement that Maxwell's electrodynamics in vacuum is already
covariant under Lorentz transformations is commonplace in the
literature. We analyse the actual meaning of that statement and
demonstrate that Maxwell's equations are perfectly fit to be
Lorentz-covariant; they become Lorentz-covariant if we construct to
be so, by postulating certain transformation properties of field
functions. In Aristotelian terms, the covariance is a plain
potentiality, but not necessarily entelechy.
\end{abstract}

\section{Introduction}
Lorentz-covariance of Maxwell's equations is certainly the key link
between classical electrodynamics and special relativity. While
there is a clear consensus in the literature that `the
electrodynamic foundation of Maxwell--Lorentz's theory is in
agreement with the principle of relativity,' and thus that Maxwell's
equations are Lorentz-covariant, the true meaning of that statement
appears to be somewhat elusive. Generally, it is demonstrated that
Maxwell's equations are Lorentz-covariant if and only if the
electric and magnetic fields and charge and current densities
appearing in them transform according to some specific
transformation laws. As is well known, this can be done basically in
two ways: either transforming directly Maxwell's equations (`steep
and difficult mountaineer's path') as Einstein originally did
\cite{AE,ROS,SCH,RR}, or employing the powerful and elegant, almost
dazzling, tensorial approach in Minkowski space-time. Neither way is
very transparent to the student.

On the other hand, the student of relativity encounters frequently
some potentially confusing locutions on Lorentz-covariance of
Maxwell's equations which, in the long run, might lead the student
to think that `requirement of form--invariance is automatically
fulfilled for Maxwell's fundamental equations of electrodynamics
{\it in vacuo}.' For example, in his classic book, M{\o}ller
\cite{MOLL} states: `we saw that it is necessary to change the
fundamental equations of mechanics in order to bring them into
accordance with the principle of relativity. This is not so with the
equations of electrodynamics in vacuum, the Maxwell equations,
which, as we shall see, are already covariant under Lorentz
transformations [...].' In the same vein, Rindler \cite{WRSR}
writes: `Having examined and relativistically modified Newtonian
particle mechanics, it would be natural to look next with the same
intentions at Maxwell's electrodynamics, at first in vacuum. But
that theory turns out to be already ``special-relativistic''. In
other words, its basic laws, as summarized by the four Maxwell
equations plus Lorentz's force law, are form-invariant under Lorentz
transformations, i. e. under transformations from one inertial frame
to another.' Similarly, Mario Bunge \cite{MB} asserts that
relativistic electrodynamics `is not a new theory but a
reformulation of CEM [classical electromagnetism], which was
relativistic without knowing it.' Also, in his fine book \cite{VAU},
Ugarov affirms: `It is remarkable that the system of Maxwell's
equations formulated fifty years prior to the advent of the special
theory of relativity proved to be covariant with respect to the
Lorentz transformation, i.e. it retains its appearance, with the
accuracy of variables' designations, under the Lorentz
transformation. This signifies that the system of Maxwell's
equations retains its appearance in any inertial frame of reference,
and the principle of relativity holds automatically.' As the last
characteristic example, I quote from a recent book by
Christodoulides \cite{CC}: `It is obvious that electromagnetic
theory, as expressed by Maxwell's equations, is a relativistic
theory, whose equations needed no modification in order to become
compatible with the Theory of Relativity, at least as these apply to
the vacuum.'

Recently, I pointed out that the above statements should be taken
{\it cum grano salis}: Lorentz-covariance of Maxwell's equations is
{\it not} fulfilled automatically \cite{DVR}. I noted that, for
example, the so-called source-free Maxwell's equations,
$\mbox{curl}\bi E = -\partial \bi B/\partial t$ and $\mbox{div}\bi B
= 0$, are Lorentz-covariant if one {\it defines} $\bi E'$ and $\bi
B'$ via $\bi E$ and $\bi B$ as given by the well-known
transformation rules. A complete but succinct discussion of the
issue is given in \cite{DVR1}. However, taking into account a
possible  relevance of the issue for the student of relativity, it
is perhaps worthwhile to discuss in some detail what the
above-mentioned authors actually meant by `Maxwell's equations are
already covariant under Lorentz transformations.'

\section{Lorentz-covariance of Maxwell's equations}

\subsection{Mathematical prelude}
The problem of Lorentz-covariance of Maxwell's equations is
basically a mathematical question. In this Subsection, for the
convenience of the reader, we recall some familiar results in the
simple way, from vectorial perspective.

We begin by writing a set of coupled partial differential equations

\begin {equation}
\frac {\partial E_z}{\partial y} - \frac {\partial E_y}{\partial z}
= - \frac {\partial B_x}{\partial t}\, ,
\end {equation}

\begin {equation}
\frac {\partial E_x}{\partial z} - \frac {\partial E_z}{\partial x}
= - \frac {\partial B_y}{\partial t}\, ,
\end {equation}

\begin {equation}
\frac {\partial E_y}{\partial x} - \frac {\partial E_x}{\partial y}
= - \frac {\partial B_z}{\partial t}\, ,
\end {equation}

\begin {equation}
\frac {\partial B_x}{\partial x} + \frac {\partial B_y}{\partial y}
+ \frac {\partial B_z}{\partial z} = 0\, ,
\end {equation}
where $E_i = E_i(x,y,z,t)$ and $B_i = B_i(x,y,z,t)$, $i$ stands for
subscripts $x,y,z$, are functions of the mutually independent
variables $x,y,z$ and $t$. Introduce another set of the mutually
independent variables $x',y',z'$ and $t'$, and let them be the
following functions of $x,y,z$ and $t$
\begin{equation}
x' = \gamma(x-vt)\, , y' = y\, , z' = z\, , t' = \gamma(t -
vx/c^2)\, ,
\end{equation}
where $\gamma \equiv (1 - v^2/c^2)^{-1/2}$, $c \equiv
\sqrt{1/\epsilon_0\mu_0}$, $\epsilon_0$ and $\mu_0$ are positive
constants, and $v$ is a nonnegative constant satisfying $0 \leq v <
c$.

As is well known, expressing unprimed by primed variables in
equations (1)-(4), employing the standard procedure which involves
the chain rule for differentiation, after some manipulations one
obtains that the following primed equations apply (a detailed
derivation is found, e.g., in \cite{ROS}, Section 8.2):

\begin {equation}
\frac {\partial {\cal E}'_z}{\partial y'} - \frac {\partial {\cal
E}'_y}{\partial z'} = - \frac {\partial {\cal B}'_x}{\partial t'}\,
,
\end {equation}

\begin {equation}
\frac {\partial {\cal E}'_x}{\partial z'} - \frac {\partial {\cal
E}'_z}{\partial x'} = - \frac {\partial {\cal B}'_y}{\partial t'}\,
,
\end {equation}

\begin {equation}
\frac {\partial {\cal E}'_y}{\partial x'} - \frac {\partial {\cal
E}'_x}{\partial y'} = - \frac {\partial {\cal B}'_z}{\partial t'}\,
,
\end {equation}

\begin {equation}
\frac {\partial {\cal B}'_x}{\partial x'} + \frac {\partial {\cal
B}'_y}{\partial y'} + \frac {\partial {\cal B}'_z}{\partial z'} =
0\, ,
\end {equation}
where

\begin{equation}
\left.
\begin{array}{ll}
{\cal E}'_x \equiv E_x                 &   \qquad \qquad {\cal B}'_x   \equiv B_x \\
{\cal E}'_y \equiv\gamma(E_y - vB_z)   &   \qquad \qquad {\cal B}'_y  \equiv \gamma(B_y +\frac {v}{c^2}E_z)\\
{\cal E}'_z \equiv\gamma(E_z + vB_y)   &  \qquad \qquad {\cal B}'_z
\equiv\gamma(B_z-\frac{v}{c^2}E_y)
\end{array}
\quad \right\}
\end{equation}
In equations (10) ${\cal E}'_i = {\cal E}'_i(x',y',z',t')$ and $E_i
= E_i[\gamma(x' + vt'),y',z',\gamma(t' + vx'/c^2)]$ and analogously
for ${\cal B}'_i$ and $B_i$. Obviously, equations (6)-(9) have the
same form as equations (1)-(4). Thus, transforming equations (1)-(4)
by transformation of variables (5), one reveals that those equations
imply that, in the primed variables, equations (6)-(9) of the same
form apply under the proviso that ${\cal E}'_i$ and ${\cal B}'_i$
therein be given by identities (10). Consequently, if $E_i$ and
$B_i$ satisfy unprimed equations (1)-(4), one knows that ${\cal
E}'_i$ and ${\cal B}'_i$ determined by identities (10) satisfy
primed equations (6)-(9).

Note that, from equations (10) and (5), {\it mutatis mutandis}, one
obtains the following inverse identities:

\begin{equation}
\left.
\begin{array}{ll}
E_x \equiv {\cal E}'_x                 &   \qquad \qquad B_x   \equiv {\cal B}'_x \\
E_y \equiv\gamma({\cal E}'_y + v{\cal B}'_z)   &   \qquad \qquad B_y  \equiv \gamma({\cal B}'_y -\frac {v}{c^2}{\cal E}'_z)\\
E_z \equiv\gamma({\cal E}'_z - v{\cal B}'_y)   &  \qquad \qquad B_z
\equiv\gamma({\cal B}'_z+\frac{v}{c^2}{\cal E}'_y)
\end{array}
\quad \right\}
\end{equation}
which of course are obtained quickly by interchanging primed and
unprimed quantities and replacing $v$ by $-v$ in (10).

Assume now that functions $E_i$ and $B_i$, in addition to equations
(1)-(4), must also satisfy another set of equations:

\begin {equation}
\frac {\partial B_z}{\partial y} - \frac {\partial B_y}{\partial z}
= \mu_0 \rho u_x + \frac{1}{c^2} \frac {\partial E_x}{\partial t}\,
,
\end {equation}

\begin {equation}
\frac {\partial B_x}{\partial z} - \frac {\partial B_z}{\partial x}
= \mu_0 \rho u_y + \frac{1}{c^2} \frac {\partial E_y}{\partial t}\,
,
\end {equation}

\begin {equation}
\frac {\partial B_y}{\partial x} - \frac {\partial B_x}{\partial y}
= \mu_0 \rho u_z + \frac{1}{c^2} \frac {\partial E_z}{\partial t}\,
,
\end {equation}

\begin {equation}
\rho = \epsilon_0 \left(\frac {\partial E_x}{\partial x} + \frac
{\partial E_y}{\partial y} + \frac {\partial E_z}{\partial z}\right)
\equiv \varrho_{\scriptscriptstyle E} \, ,
\end {equation}
where $\rho(x,y,z,t)$ is the `charge density' (without attaching any
physical meaning to it), and $u_i = u_i(x,y,z,t)$ are Cartesian
components of the {\it velocity field} of the `charge.' The only
constraint imposed by equations (12)-(15) on $\rho$ and $u_i$ is the
`equation of continuity,'

\begin {equation}
\frac {\partial (\rho u_x)}{\partial x} + \frac {\partial (\rho
u_y)}{\partial y} + \frac {\partial (\rho u_z)}{\partial z} + \frac
{\partial \rho}{\partial t} = 0\, ,
\end {equation}
which is a necessary condition for the validity of equations
(12)-(15). Thus, the equation of continuity may apply even if
equations (12)-(15) do not apply \cite{JAH}.

Introduce symbol
\begin {equation}
\varrho'_{\scriptscriptstyle E} \equiv \epsilon_0 \left(\frac
{\partial {\cal E}'_x}{\partial x'} + \frac {\partial {\cal
E}'_y}{\partial y'} + \frac {\partial {\cal E}'_z}{\partial
z'}\right) \, ,
\end {equation}
where ${\cal E}'_i$ are given by identities (10). Employing the
standard procedure, one finds that $\varrho'_{\scriptscriptstyle E}$
transforms according to equation

\begin {equation}
\varrho'_{\scriptscriptstyle E} = \gamma
\left[\varrho_{\scriptscriptstyle E} -\epsilon_0 v \left(\frac
{\partial B_z}{\partial y} - \frac {\partial B_y}{\partial z} -
\frac{1}{c^2} \frac {\partial E_x}{\partial t}\right)\right]\, ,
\end {equation}
wherefrom using equations (12) and (15) one obtains

\begin {equation}
\varrho'_{\scriptscriptstyle E} = \gamma \left(\rho  - \frac
{v}{c^2}\rho u_x\right)\, ,
\end {equation}

Making use of the familiar transformations for velocity field
components,

\begin {equation}
u'_x = \frac {u_x - v}{1 - u_xv/c^2}\, ,  \quad u'_y = \frac
{u_y}{\gamma(1 - u_xv/c^2)}\, , \quad u'_z = \frac {u_z}{\gamma(1 -
u_xv/c^2)}\, ,
\end {equation}
and their inverse, from eq. (19) one gets

\begin {equation}
\rho = \gamma \left(\varrho'_{\scriptscriptstyle E} + \frac
{v}{c^2}\varrho'_{\scriptscriptstyle E}u'_x\right) \, .
\end {equation}

Transforming eq. (12) through eqs. (11) yields directly

\begin {equation}
\frac {\partial {\cal B}'_z}{\partial y'} - \frac {\partial {\cal
B}'_y}{\partial z'} = \mu_0 \left(\frac {\rho u_x}{\gamma} -
\varrho'_{\scriptscriptstyle E}v\right) + \frac{1}{c^2} \frac
{\partial {\cal E}'_x}{\partial t'}\, ,
\end {equation}
which using eq. (19) and the first formula (20) gives

\begin {equation}
\frac {\partial {\cal B}'_z}{\partial y'} - \frac {\partial {\cal
B}'_y}{\partial z'} = \mu_0 \left[\gamma\left(\rho - \frac {v
}{c^2}\rho u_x\right)\right]u'_x + \frac{1}{c^2} \frac {\partial
{\cal E}'_x}{\partial t'}\, ,
\end {equation}

Transforming in the same way eqs. (13)-(15), and using formulas (20)
and eq. (23), one gets

\begin {equation}
\frac {\partial {\cal B}'_x}{\partial z'} - \frac {\partial {\cal
B}'_z}{\partial x'} = \mu_0 \left[\gamma\left(\rho - \frac {v
}{c^2}\rho u_x\right)\right]u'_y + \frac{1}{c^2} \frac {\partial
{\cal E}'_y}{\partial t'}\, ,
\end {equation}

\begin {equation}
\frac {\partial {\cal B}'_y}{\partial x'} - \frac {\partial {\cal
B}'_x}{\partial y'} = \mu_0 \left[\gamma\left(\rho - \frac {v
}{c^2}\rho u_x\right)\right]u'_z + \frac{1}{c^2} \frac {\partial
{\cal E}'_z}{\partial t'}\, ,
\end {equation}

\begin {equation}
\gamma  \left(\rho - \frac {v }{c^2}\rho u_x\right) =
\varrho'_{\scriptscriptstyle E} \, ,
\end {equation}
respectively. Eq. (26) is identical with eq. (19), as it should be.

Equations (1)-(4) and (12)-(15) can obviously be recast into the
compact form

\begin{equation}
\fl \bnabla \times\bi E = -\frac {\partial \bi B}{\partial t}\,
,\bnabla \cdot \bi B = 0\, ,\bnabla \times\bi B =\mu_0 \rho \bi u +
\epsilon_0\mu_0 \frac {\partial \bi E}{\partial t}\, ,\bnabla \cdot
\bi E = \frac {\rho}{\epsilon_0}\, ,
\end{equation}
and the transformed equations (6)-(9) and (23)-(26) can be recast
into
\begin{equation}
\fl \bnabla' \times\pmb{\cal E}' = -\frac {\partial \pmb{\cal
B}'}{\partial t'}\, ,\bnabla' \cdot \pmb{\cal B}' = 0\, , \bnabla'
\times\pmb{\cal B}' =\mu_0 \varrho' \bi u' + \epsilon_0\mu_0 \frac
{\partial \pmb{\cal E}'}{\partial t'}\, ,\bnabla' \cdot \pmb{\cal
E}' = \frac {\varrho'}{\epsilon_0}\, ,
\end{equation}
where $\varrho' \equiv \gamma  \left(\rho - \frac {v }{c^2}\rho
u_x\right)$.

Thus, transforming equations (27) by transformation of variables
(5), one obtains that, in primed variables, equations (28) of the
same form apply under the proviso that $\pmb{\cal E}'$ and
$\pmb{\cal B}'$ therein be defined by identities (10). Consequently,
if $\bi E$ and $\bi B$ satisfy unprimed equations (27), one knows
that $\pmb{\cal E}'$ and $\pmb{\cal B}'$ defined by identities (10)
satisfy primed equations (28). This is all one can extract from the
unprimed Maxwell's equations (27), transforming them by the Lorentz
transformation (5).

\subsection{Are Maxwell's equations Lorentz-covariant?}

Comparing equations (27) and (28), the following conclusion is
readily reached: in order that Maxwell's equations which apply in
unprimed variables, hold also in primed variables, (that is, in
standard parlance, in order that Maxwell's equations be
Lorentz-covariant), it suffices to {\it define} $\bi E'$ and $\bi
B'$ (by which we mean {\it counterparts} of $\bi E$ and $\bi B$ in
primed variables) by equations
\begin{equation}
\bi E' \stackrel{d}{=} \pmb{\cal E}'\, ,\qquad \bi B'
\stackrel{d}{=} \pmb{\cal B}'\, ,
\end{equation}
{\it and} the `charge density' in primed variables, $\rho'$, by
equation

\begin {equation}
\rho' \stackrel{d}{=} \gamma  \left(\rho- \frac {v}{c^2} \rho
u_x\right) \, .
\end {equation}
Since transformation properties of $\bi E$, $\bi B$ and $\rho$ are
not known beforehand, $\bi E'$, $\bi B'$ and $\rho'$ have to be {\it
defined} by equations (29) and (30), i.e. through transformation
laws.\footnote[1] {While it appears that a multiplicative factor
$\Psi(v)$ could be included in eqs. (29) and (30), a simple analysis
demonstrates that $\Psi(v)$ must equal one \cite{AE}.} Introducing
those definitions `is an inconspicuous but indispensable step, a
{\it conditio sine qua non} for Lorentz-covariance of Maxwell's
equations in the strict sense of the word' \cite{DVR1}.

Now we are armed with all the facts necessary to answer our query,
are Maxwell's equations Lorentz-covariant. That is, do Maxwell's
equations retain their form under transformation of variables (5)?
The correct answer appears to be: the Maxwell equations are
ready-made to be Lorentz-covariant, but they are actually
Lorentz-covariant only if we {\it construct} to be so (cf, e.g.,
\cite{STR,FRE}). As was demonstrated above, what exactly is
sufficient to be postulated for the covariance emanates from the
equations themselves. In this sense, and in this sense only, one can
speak about `a miracle [that] Maxwell, fully unaware of relativity,
had nevertheless written his equations in a relativistically
covariant form straight away' \cite{EF}. However, the covariance is
not fulfilled automatically; there is no covariance without
postulating specific transformation properties of the quantities
appearing in the equations (with the exception of course of purely
geometric quantities, such as velocity and acceleration, which are
already defined in both unprimed and primed coordinates, and whose
transformation properties follow from the definitions). In
Aristotelian terms, Lorentz-covariance is contained in Maxwell's
equations as a plain potentiality, but not as entelechy. One should
keep this in mind.\footnote[2] {Incidentally, recall that the
equation of continuity (16) {\it itself} is ready-made to be not
only Lorentz-covariant but also Galilei-covariant (cf, e.g.,
\cite{DVR2,CW}). Whichever covariance is preferred on physical
grounds, the remaining one then becomes a purely mathematical
property.}

Einstein's original demonstration that `the electrodynamic
foundation of Lorentz's theory of the electrodynamics of moving
bodies agrees with the principle of relativity,' \cite{AE} is
basically mathematics disguised as physics. Einstein postulates that
Maxwell's  equations conform to the principle of relativity and thus
that both eqs. (27) and equations

\begin{equation}
\fl \bnabla' \times\bi E' = -\frac {\partial \bi B'}{\partial t'}\,
,\bnabla' \cdot \bi B' = 0\, ,\bnabla' \times\bi B' =\mu_0 \rho' \bi
u' + \epsilon_0\mu_0 \frac {\partial \bi E'}{\partial t'}\,
,\bnabla' \cdot \bi E' = \frac {\rho'}{\epsilon_0}\, ,
\end{equation}
hold; basically, he thus postulates that Maxwell's equations are
Lorentz--covariant. Comparing (mutually equivalent) equations (28)
and (31), he deduces the necessary conditions for the covariance
(eqs. (29) and (30) regarded as transformation laws).

\section{Concluding comments}
Take now that the Lorentz transformation (5) has its received
physical meaning, i.e., assume that it relates space and time
coordinates of an event in a given inertial frame $S$ with the space
and time coordinates of the same event in an inertial frame $S'$
which is in a standard configuration with $S$. As is well known,
assuming the validity of Maxwell's equations in the given frame $S$,
and also taking that $\bi E'$, $\bi B'$ and $\rho'$ are {\it
defined} by equations (29) and (30) (achieving thus
Lorentz-covariance of Maxwell's equations), would ensure the
validity of Maxwell's equations in any reference frame $S'$ in
uniform translation with respect to $S$, if the special theory of
relativity is valid. In this context, definitions (29) and (30)
express the electric and magnetic fields and charge density in $S'$,
and thus, basically, represent a fundamental physical assumption.
However, as Bartocci and Mamone Capria pointed out, the plain {\it
possibility} of achieving Lorentz-covariance of Maxwell's equations
can be regarded as nothing more than an interesting mathematical
property devoid of any physical contents \cite{UB}. It is perhaps
instructive to recognize that the {\it formal} covariance can be
employed as a handy tool, quite outside the relativistic framework
\cite{DVR}.

To summarize, in the context of physics, if Maxwell's equations
describe physical fields in an inertial frame $S$, and Lorentz
transformations relate space and time coordinates of the same event
as observed in two inertial frames $S$ and $S'$ in relative motion,
Lorentz-covariance of Maxwell's equations expresses a fundamental
physical assumption that the same (primed!) Maxwell's equations
describe the physical fields also in the $S'$ frame. On the other
hand, from the mathematical side, what is latent in Maxwell's
equations is, first, that they are ready-made to be
Lorentz-covariant, and, second, the precise `recipe' how to achieve
that they actually be Lorentz-covariant. Shortly, Maxwell's
equations are Lorentz-covariant if we construct to be so, but they
need not be.\footnote[4] {Thus, Rindler's formulation that Maxwell's
equations `fit {\it perfectly} into the scheme of special
relativity' \cite{WRR}, should perhaps be amended as `{\it can} fit
{\it perfectly} into the scheme of special relativity.'} However, it
was indeed a miracle that Maxwell had written his equations in a
form perfectly fit to be Lorentz-covariant. From this perspective,
Heinrich Hertz's {\it feeling} that Maxwell's equations `give back
to us more than was originally put into them,' proved prophetic.

Finally, note that, as is well known, analysis of Maxwell's
equations can be often made much easier in terms of potentials. For
the sake of completeness, a brief discussion of Lorentz-covariance
of Maxwell's equations from the perspective of potentials, skipping
the familiar details, is given in Appendix.

\section*{Acknowledgments}
I thank Vladimir Onoochin for stimulating correspondence. My work is
supported by the Ministry of Science and Education of the Republic
of Serbia, project No. 451-03-47/2023-01/200162.

\appendix
\section*{Appendix}
\setcounter{section}{1}

First of all, recall that Maxwell's equations imply the equation of
continuity (16):

\begin {equation}
 \frac {\partial \rho}{\partial t} + \frac {\partial (\rho u_x)}{\partial x}
+ \frac {\partial (\rho u_y)}{\partial y} + \frac {\partial (\rho
u_z)}{\partial z} = 0\, ,
\end {equation}
which is a necessary condition for the validity of Maxwell's
equations and thus it may be valid even if Maxwell's equations do
not apply. Recall also that Maxwell's equations possess, {\it inter
alia}, a nice property that they allow themselves to be considerably
simplified mathematically by expressing $E_i$ and $B_i$ in terms of
potentials $\Phi$ and $A_i$ introduced by

\begin{equation}
\left.
\begin{array}{ll}
E_x = -  \frac {\partial \Phi}{\partial x} -  \frac {\partial A_x}{\partial t}   &   \qquad B_x = \frac {\partial A_z}{\partial y} -  \frac {\partial A_y}{\partial z}  \\

E_y = -  \frac {\partial \Phi}{\partial y} -  \frac {\partial A_y}{\partial t}   &   \qquad B_y = \frac {\partial A_x}{\partial z} -  \frac {\partial A_z}{\partial x}  \\

E_z = -  \frac {\partial \Phi}{\partial z} -  \frac {\partial
A_z}{\partial t}   & \qquad B_z = \frac {\partial A_y}{\partial x} -
\frac {\partial A_x}{\partial y}

\end{array}
\quad \right\}
\end{equation}
Assuming that the potentials satisfy the Lorenz gauge condition,

\begin {equation}
\frac {1}{c^2}\frac {\partial \Phi}{\partial t} + \frac {\partial
A_x}{\partial x} + \frac {\partial A_y}{\partial y} + \frac
{\partial A_z}{\partial z} = 0\, ,
\end {equation}
it follows that the potentials then should satisfy the inhomogeneous
d'Alembert type equations:

\begin {equation}
\Box \Phi = -\frac {\rho}{\epsilon_0}\, ,
\end {equation}

\begin {equation}
\Box A_x = -\mu_0\rho u_x\, , \quad \Box A_y = -\mu_0\rho u_y\, ,
\quad \Box A_z = -\mu_0\rho u_z\, ,
\end {equation}
where
\begin {equation}
\Box \equiv  \frac {\partial^2 }{\partial x^2} + \frac {\partial^2
}{\partial y^2} + \frac {\partial^2 }{\partial z^2} -  \frac
{1}{c^2}\frac {\partial^2 }{\partial t^2}\, ,
\end {equation}
Now transform the continuity equation (A.1) replacing unprimed by
primed variables according to equations (5), employing formulae for
changing partial differential coefficients
\begin {equation}
\fl  \frac {\partial }{\partial t} = \gamma \left(\frac {\partial
}{\partial t'} - v\frac {\partial }{\partial x'}\right)\, \quad
\frac {\partial }{\partial x} = \gamma \left(\frac {\partial
}{\partial x'} - \frac {v}{c^2}\frac {\partial }{\partial t'}\right)
\, , \quad \frac {\partial }{\partial y} = \frac {\partial
}{\partial y'}\, , \quad \frac {\partial }{\partial z} = \frac
{\partial }{\partial z'}\, .
\end {equation}
One obtains
\begin {equation}
\fl \frac {\partial }{\partial t'}\gamma (\rho - \frac {v }{c^2}\rho
u_x) + \frac {\partial }{\partial x'}\gamma (\rho u_x - \rho v) +
\frac {\partial }{\partial y'}(\rho u_y) + \frac {\partial
}{\partial z'}(\rho u_z) = 0\, .
\end {equation}
Using equations (20), one has

\begin {equation}
\fl \frac {\partial }{\partial t'}\gamma (\rho - \frac {v }{c^2}\rho
u_x) + \frac {\partial }{\partial x'}\gamma \rho u'_x(1 - \frac
{u_xv }{c^2}) + \frac {\partial }{\partial y'}\gamma \rho u'_y(1 -
\frac {u_xv }{c^2}) + \frac {\partial }{\partial z'}\gamma \rho
u'_z(1 - \frac {u_xv }{c^2}) = 0\, .
\end {equation}
Inspecting the last equation, the following conclusion is readily
reached: in order that equation of continuity (A.1) implies equation
of the same form and content in primed variables, it suffices to
{\it define} the charge density in primed coordinates, $\rho'$, by

\begin {equation}
\rho' = \gamma  \left(\rho- \frac {v}{c^2} \rho u_x\right) \, .
\end {equation}
With that definition, equation (A.9) obviously reduces to

\begin {equation}
 \frac {\partial \rho'}{\partial t'} + \frac {\partial (\rho' u'_x)}{\partial x'}
+ \frac {\partial (\rho' u'_y)}{\partial y'} + \frac {\partial
(\rho' u'_z)}{\partial z'} = 0\, ,
\end {equation}
which is identical with equation (A.1), except for primes. Functions
$\rho' u'_i$ are Cartesian components of the convection current
density in the $S'$ frame, as $\rho u_i$ are in the $S$ frame.
Clearly, $\rho c$ , $\rho u_x$, $\rho u_y$ and $\rho u_z$ transform
according to the rules
\begin {equation}
\fl \rho'c = \gamma (\rho c - \frac {v }{c}\rho u_x)\, , \quad
\rho'u'_x = \gamma (\rho u_x - \frac {v }{c}\rho c)\, , \quad
\rho'u'_y = \rho u_y\, , \quad \rho'u'_z = \rho u_z\, ,
\end {equation}
under the Lorentz transformation (5).

Now transform in the same way another simple equation, the Lorenz
gauge condition (A.3). One obtains automatically
\begin {equation}
\fl  \frac {1}{c^2}\frac {\partial }{\partial t'}\gamma (\Phi -
vA_x) + \frac {\partial }{\partial x'}\gamma (A_x - \frac {v
}{c^2}\Phi) +  \frac {\partial }{\partial y'}A_y + \frac {\partial
}{\partial z'}A_z = 0\, .
\end {equation}
Obviously, in order that equation (A.3) be Lorentz-covariant, it
suffices to {\it define} functions of primed variables $\Phi'$,
$A'_x$, $A'_y$ and $A'_z$ by equations

\begin {equation}
\fl  \Phi' = \gamma (\Phi - vA_x)\, , \quad A'_x = \gamma (A_x -
\frac {v }{c^2}\Phi)\, , \quad  A'_y = A_y\, , \quad A'_z = A_z\, ,
\end {equation}
the result that Poincar\' e reached a long time ago by a different
path, postulating charge invariance and invariance of the
inhomogeneous d'Alembert type equations for the potentials
\cite{HP}. Since the primed functions are {\it defined} by equations
(A.14), it follows that $\Phi$, $A_x$, $A_y$ and $A_z$ {\it a
fortiori} transform according to equations (A.14) under the Lorentz
transformation (5). For convenience, recast the transformation rules
into

\begin {equation}
\fl  \frac {\Phi'}{c} = \gamma \left( \frac {\Phi}{c}  - \frac
{v}{c}A_x\right )\, , \quad A'_x = \gamma \left(A_x - \frac {v
}{c}\frac {\Phi}{c}\right)\, , \quad  A'_y = A_y\, , \quad A'_z =
A_z\, .
\end {equation}

A glance at equations (A.12) reveals that, with {\it definition}
(A.10) of $\rho'$ density, $\rho c$, $\rho u_x$, $\rho u_y$ and
$\rho u_z$ become contravariant components of a 4-vector of
Minkowski space-time; equation (A.15) shows that the analogous
conclusion applies to $\Phi/c$, $A_x$, $A_y$ and $A_z$. Thus,
4-current density $J^\mu$ and 4-potential $A^\mu$ are constructed.

Now we arrived at the familiar, wide and well trodden path, and no
need to go further. Namely, as is well known, in the tensorial
notation, with 4-vectors $J^\mu$ and $A^\mu$, Lorentz-covariance of
Maxwell's equations is an obvious fact, offered as on a plate. What
is perhaps less obvious is that, instead of simply asserting that
$\Phi$ and $\bi A$ together constitute a 4-vector, it would be more
correct to specify that now $\Phi$ and $\bi A$ together constitute a
4-vector {\it per definitionem}, namely, we constructed to be
so.\footnote[5] {As Rindler (\cite{WRSR}, p 155) notes, `[...] we
can construct a tensor by specifying its components arbitrarily in
{\it one} coordinate system, say $\{x^i\}$, and then using the
transformation law [expressing the familiar informal definition of
tensors] to define its components in all other systems, or, in the
case of a qualified tensor, in all those systems which are mutually
connected by transformations belonging to the chosen subgroup.' }
The same remark applies to $\rho$ and $\rho\bi u$. Of course, in the
latter case, the transformation rules (A.12) can be obtained without
construction, as a consequence of the principle of charge
invariance.

Thus, in the language of potentials and 4-tensors, our main
conclusion is reached in a simpler and more transparent way.
Maxwell's equations are perfectly fit to be Lorentz-covariant; they
become Lorentz-covariant only if we define the primed potentials and
charge density so that $(\Phi/c, \bi A)$ and $(\rho c, \rho \bi u)$
be 4-vectors of Minkowski space-time. Deciding that $(\Phi/c, \bi
A)$ and $(\rho c, \rho \bi u)$ be contravariant components of
4-vectors ensures Lorentz-covariance of Maxwell's equations,
enabling us to recast those equations in an explicitly
Lorentz-covariant form.

Thus, we can agree with Sommerfeld's \cite {SOM} simile that `the
true mathematical structure of these entities [$\Phi$ and $\bi A$]
will appear only now [in the language of 4-tensors], as in a
mountain landscape when the fog lifts,' only in the framework of the
interpretation given above. Sommerfeld's claim that `by reducing the
Maxwell equations to the four-vector [$A^{\mu}$, $J^{\mu}$ and the
operators $\partial_{\mu}$, $\Box = - \partial_{\mu}\partial^{\mu}$]
we have demonstrated at the same time their {\it general validity},
independent of the coordinate system' is, strictly speaking,
incorrect. Basically, we have {\it assumed} a four-vector character
of $(\Phi/c, \bi A)$ and $(\rho c, \rho \bi u)$ and thus we have
constructed that `the Maxwell equations satisfy the relativity
postulate from the very beginning.' This constructional aspect of
Lorentz-covariance of Maxwell's equations, clearly enunciated by
Einstein \cite {AE2} and by Bergmann \cite{PGB}, seems to be
understated in the literature.

\end{document}